\documentclass[11pt]{article}
\usepackage{hyperref}
\pdfoutput=1

\begin{document}

\title{Vertically-Vibrated Gas-Liquid Interfaces: \\ Surface Deformation and Breakup}

\author{T. J. O'Hern, E. F. Romero, C. F. Brooks, B. Shelden, \and J. R. Torczynski, A. M. Kraynik, L. A. Romero, G. L. Benavides \\
\\\vspace{6pt} Sandia National Laboratories \\ Albuquerque, NM 87185, USA}

\maketitle

In his pioneering work of 1831, Michael Faraday\footnotemark \, demonstrated that a vertically vibrated gas-liquid interface exhibits a period-doubling 
bifurcation from a flat state to a wavy configuration at certain frequencies or amplitudes. Typical experiments performed using thin layers 
of water produce "Faraday ripples", modest-amplitude nonlinear standing waves. However, experiments by Hashimoto and Sudo\footnotemark \, and Jameson\footnotemark \, as well as those performed 
in the present study show that 
much more dramatic disturbances can be generated at the gas-liquid free surface under certain ranges of vibration conditions.
A series of fluid dynamics videos have been recorded to demonstrate some of the phenomenology that occur when gas-liquid interfaces are 
forced to break up by vertical vibrations within critical ranges of frequency and amplitude. 
This breakup mechanism was examined experimentally using deep layers of polydimethylsiloxane (PDMS) silicone oils over a range of viscosity 
and sinusoidal, 
primarily axial vibration conditions that can produce dramatic disturbances at the gas-liquid free surface. Although small-amplitude 
vibrations produce standing Faraday waves, large-amplitude vibrations produce liquid jets into the gas, droplets pinching off from 
the jets, gas cavities in the liquid from droplet impact, and bubble transport below the interface, all of which can be seen in the videos. Experiments were performed using 
several different PDMS silicone oils covering a range of viscosity (1-50 cSt, i.e., 1-50 times nominal water) and relatively low surface
tension (e.g., 0.25 times nominal water) over a range of vibration conditions (100-150 Hz, less than 300 micron displacement, yielding 
accelerations ranging from 4 to 7 times gravitational acceleration). 
The cylinder containing the silicone oil was 25.4 mm inner diameter
and approximately 100 mm long. It was typically filled halfway with the PDMS liquid, with ambient air above. High-speed videos were recorded using 
Phantom V. 9.1 cameras running at 600 to 1000 frames per second. Applications include liquid fuel rockets, inertial sensing devices, 
moving vehicles, mixing processes, and acoustic excitation.

The videos showing the surface deformation and breakup under various conditions can be seen in the ancillary files. This video has been submitted to the American Physical Society Division of Fluid Dynamics (APS DFD) Gallery of Fluid Motion 2010 which is an 
annual showcase of fluid dynamics videos.

\footnotetext[1]{Faraday, M., On the forms and states assumed by fluids in contact with vibrating elastic surfaces, Philosophical Transactions of the Royal Society of London, Vol. 121, pp. 319-340, 1831.}

\footnotetext[2]{Hashimoto, H. and Sudo, S., Surface Disintegration and Bubble Formation in Vertically Vibrated Liquid Column, AIAA Journal, Vol. 18, No. 4, pp. 442-449, 1980.}

\footnotetext[3]{Jameson, G. J., The motion of a bubble in a vertically oscillating viscous liquid, Chemical Engineering Science, Vol. 21, pp. 35-38, 1966.}

Sandia is a multiprogram laboratory operated by Sandia Corporation, a 
Lockheed Martin Company, for the United States Department of Energy's National Nuclear Security Administration under contract DE-AC04-94AL85000. 

\end{document}